\documentclass [12pt]{article}
\usepackage{axodraw}

\catcode`@=12
\newcommand{\be}{\begin{equation}}
\newcommand{\ee}{\end{equation}}
\newcommand{\bea}{\begin{eqnarray}}
\newcommand{\eea}{\end{eqnarray}}

\begin{document}
\thispagestyle{empty}
\begin{flushright} UCRHEP-T331\\February 2002
\end{flushright}
\vspace{0.5in}
\begin{center}
{\LARGE \bf Neutrinoless Double Beta Decay\\
with Negligible Neutrino Mass\\}
\vspace{1.5in}

{\bf Biswajoy Brahmachari$^{a}$ and Ernest Ma$^{b}$\\}

\vskip 1cm

(a) Theoretical Physics Group, Saha Institute of Nuclear Physics,\\
AF/1 Bidhannagar, Kolkata (Calcutta) 700064, INDIA \\
\vskip .5cm

(b) Physics Department, University of California,\\ 
Riverside, California, CA 92521, USA\\
\vskip .5cm

\end{center}

\vskip 1.5in

\begin{center}
\underbar{Abstract} \\
\end{center}

If the electron neutrino has an effective nonzero Majorana mass, then 
neutrinoless double beta decay will occur.  However, the latter is possible 
also with a negligible neutrino mass.  We show how this may happen in 
a simple model of scalar diquarks and dileptons.  This possibility allows 
neutrino masses to be small and hierarchical, without conflicting with 
the possible experimental evidence of neutrinoless double beta decay.

\newpage
\baselineskip 24pt

With the established evidence of atmospheric \cite{atm} and solar \cite{solar} 
neutrino oscillations, the notion of neutrino mass is generally accepted. 
Whereas there is yet no direct evidence of neutrino mass in beta decay 
\cite{beta}, there is now a report \cite{klapdor} of the first positive 
evidence of neutrinoless double beta decay \cite{double}, which is commonly 
interpreted as being due to an effective nonzero Majorana mass of the electron 
neutrino.  With this assumption, one may then explore the consequences 
\cite{recent} of having neutrino masses constrained by oscillations as well 
as neutrinoless double beta decay.

On the other hand, it is theoretically possible to have measurable 
neutrinoless double beta decay without having a corresponding Majorana 
neutrino mass of the expected magnitude \cite{rev}.  In other words, the 
mechanism responsible for neutrinoless double beta decay may generate only a 
negligible Majorana neutrino mass. [Since lepton number is violated by two 
units, a nonzero Majorana neutrino mass is unavoidable, but it may be very 
small.] 

Different mechanisms have been proposed in the past for contributions to 
neutrinoless double beta decay other than the electron neutrino Majorana 
mass.  One possibility is to add a Higgs triplet $(h^{++}, h^+, h^0)$ in 
an $SU(2)_L \times U(1)_Y$ theory \cite{higgs1}.  The couplings 
$W^- W^- h^{++}$ and $h^{++}e^- e^-$ together can produce a quark-level 
diagram mimicking neutrinoless double beta decay.  A second way is to embed 
the standard model in a left-right symmetric model \cite{rhn}.  In that case, 
$W_R$ exchange can also give rise to neutrinoless double beta decay. 
A combination of the right-handed neutrino Majorana mass and the $W_R$ 
mass scale is then constrained by experiment.  In supersymmetric theories, 
the exchange of scalar quarks can mediate neutrinoless double beta decay 
if $R$-parity (lepton parity) is violated \cite{rpty}.  In this case, the 
diagram must include two $\lambda^\prime$ vertices, where each vertex 
violates lepton number by one unit.  Vector-scalar contributions to 
neutrinoless double beta decay are also possible \cite{bm}.  The 
lepton-number-violating vertex  $u e^- \tilde{b}^c$ in supersymmetry 
together with $u d^c W^-$ can induce neutrinoless double beta decay, 
whereas the vertices $u d^c \phi^-$ and $u d^c W^-$ may do the same in a 
left-right symmetric theory, with $\phi^-$ coming from the (2,2,0) 
representation.

Here we propose instead a simple model of scalar diquarks and dileptons 
\cite{bilinear} which has the following properties.

(1) Neutrinoless double beta decay occurs through the trilinear coupling of 
2 scalar diquarks and 1 scalar dilepton.

(2) The Majorana neutrino mass generated by the above trilinear coupling 
occurs only in 4 loops and is negligibly small. [The dominant contributions 
to neutrino mass are assumed to come from some other mechanism not related 
to that of neutrinoless double beta decay.]

(3) The proposed scalar diquarks and dileptons interact only with 
first-generation fermions, i.e. $u$, $d$, and $e$.  [This avoids constraints 
from $\mu \to e e e$ and $K^0 - \overline {K^0}$ mixing, etc.]

(4) The smallness of $m_u$, $m_d$, and $m_e$ is understood in terms of a 
simple mechanism \cite{small} in the Higgs sector.

(5) The model is verifiable experimentally at the TeV energy scale.

We now describe our model.  Under $SU(3)_C \times SU(2)_L \times U(1)_Y$, 
the standard-model particle content is extended to include two scalar 
diquarks
\begin{equation}
\Delta_u \sim (6,1,4/3), ~~~ \Delta_d \sim (6,1,-2/3),
\end{equation}
and one scalar dilepton
\begin{equation}
\Delta_e \sim (1,1,-2),
\end{equation}
as well as a second Higgs doublet $\Phi_2 \sim (\phi_2^+,\phi_2^0)$.  We 
assume a discrete $Z_3$ symmetry ($\omega^3 = 1$) under which
\begin{equation}
d_R, e_R, \Delta_u \sim \omega, ~~~ u_R, \Delta_d, \Delta_e, \Phi_2 \sim 
\omega^2,
\end{equation}
and all other fields $\sim 1$.  Thus the allowed Yukawa couplings are
\begin{equation}
\Delta_u^* u_R u_R, ~~ \Delta_d^* d_R d_R, ~~ \Delta_e^* e_R e_R, ~~
\overline {(u,d)}_L u_R \tilde \Phi_2, ~~ \overline {(u,d)}_L d_R \Phi_2, 
~~ \overline {(\nu,e)}_L e_R \Phi_2,
\end{equation}
where $\tilde \Phi_2 \equiv (\bar \phi_2^0, -\phi_1^-)$.  In the above, 
baryon number ($B$) and lepton number ($L$) are still conserved because 
$\Delta_u$ and $\Delta_d$ may be assigned $B = 2/3$, and $\Delta_e$ may be 
assigned $L = 2$.  However, if the $Z_3$ symmetry is now assumed to be 
broken by the explicit soft terms $\Phi_1^\dagger \Phi_2$ and $\Delta_u 
\Delta_d^* \Delta_e$, then $B$ is still conserved but $L$ is broken down to 
$(-1)^L$, i.e. lepton parity.

%%%%%%%%%%%%%%%
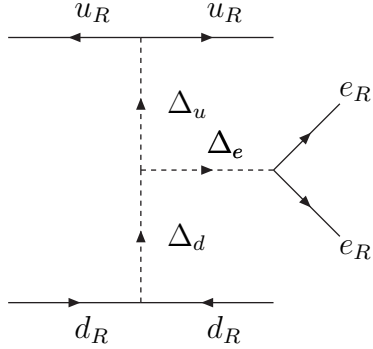
\begin{figure} % fig1
\begin{center}
%%%%%%%%%%%%%%%
\begin{picture}(200,160)(-50,30)
\ArrowLine(0,150)(-50,150)
\ArrowLine(0,150)(50,150)
\DashArrowLine(0,100)(0,150){2}
\DashArrowLine(0,100)(50,100){2}
\ArrowLine(50,100)(75,125)
\ArrowLine(50,100)(75,75)
\DashArrowLine(0,50)(0,100){2}
\ArrowLine(-50,50)(0,50)
\ArrowLine(50,50)(0,50)
\Text(-25,160)[l]{$u_R$}
\Text(25,160)[l]{$u_R$}
\Text(-25,40)[l]{$d_R$}
\Text(25,40)[l]{$d_R$}
\Text(10,125)[l]{$\Delta_u$}
\Text(10,75)[l]{$\Delta_d$}
\Text(25,110)[l]{$\Delta_e$}
\Text(75,130)[l]{$e_R$}
\Text(75,70)[l]{$e_R$}
\Text(25,110)[l]{$\Delta_e$}
\end{picture}
%%%%%%%%%%%%%%%
\end{center}
\caption[]{Diagram for neutrinoless double beta decay.
\label{fig1}
}
\end{figure} % fig1
%%%%%%%%%%%%%%%%%%%%%%%%

The $\Phi_1^\dagger \Phi_2$ term allows $\phi_2^0$ to acquire a small vacuum 
expectation value naturally \cite{small} so that the smallness of $m_u$, 
$m_d$, and $m_e$ may be understood.  The other term violates lepton 
number by two units and may be responsible for neutrinoless double beta 
decay as shown in Fig.~1.  Its amplitude is given by 
\bea
{\cal A}_{\beta \beta 0 \nu} &=& 
{h_u~h_d~h_e ~m^{soft}_{ude} \over m^2_{\Delta_u}~ m^2_{\Delta_d} 
~m^2_{\Delta_e}},
\eea
where $h_u,h_d,h_e$ are the dimensionless Yukawa couplings of ${\Delta_u},
{\Delta_d},{\Delta_e}$ with $u,d,e$ pairs and $m^{soft}_{ude}$ is a soft 
mass term.  As a crude estimate, if the effective Majorana mass of the 
electron neutrino is about 0.4 eV \cite{klapdor}, then ${\cal A}_{\beta 
\beta 0 \nu}$ should be about $10^{-16}$ GeV$^{-5}$ \cite{rev}.  This may 
be satisfied for example with $h_u = h_d = h_e = 1$, $m_{\Delta_u} = 
m_{\Delta_d} = m_{\Delta_e} = 1$ TeV, and $m^{soft}_{ude} = 100$ GeV.
From Fig.~1, it is also easy to see that its contribution to the 
electron-neutrino Majorana mass requires four loops, with two $\overline {u} 
d W^+$ vertices, two $\overline {\nu} e W^+$ vertices, and six helicity 
flips, which means that it is very much negligible.  The origin of neutrino 
mass in this model is assumed to come from some other kind of physics, 
with negligible contribution to neutrinoless double beta decay.

We have assumed that baryon number is strictly conserved, but if that 
assumption is relaxed, then the soft term $\Delta_u \Delta_d \Delta_d$ 
is allowed and neutron-antineutron oscillation becomes possible.  This 
has been discussed previously \cite{nieves}.  In our case, the discrete 
$Z_3$ symmetry also helps to suppress flavor-changing neutral currents, as 
discussed below. 

Consider the most general scalar potential of the 2 assumed scalar 
doublets \cite{small}:
\begin{eqnarray}
V &=& m_1^2 \Phi_1^\dagger \Phi_1 + m_2^2 \Phi_2^\dagger \Phi_2 + {1 \over 2} 
\lambda_1 (\Phi_1^\dagger \Phi_1)^2 + {1 \over 2} \lambda_2 (\Phi_2^\dagger 
\Phi_2)^2 \nonumber \\ && + ~\lambda_3 (\Phi_1^\dagger \Phi_1)(\Phi_2^\dagger 
\Phi_2) + \lambda_4 (\Phi_1^\dagger \Phi_2)(\Phi_2^\dagger \Phi_1) + 
[\mu_{12}^2 \Phi_1^\dagger \Phi_2 + h.c.],
\end{eqnarray}
where the $\mu_{12}^2$ term breaks the discrete $Z_3$ symmetry softly.  The 
equations of constraint for $v_{1,2} \equiv \langle \phi_{1,2}^0 \rangle$ 
are then
\begin{eqnarray}
v_1 [m_1^2 + \lambda_1 v_1^2 + (\lambda_3 + \lambda_4) v_2^2] + \mu_{12}^2 
v_2 &=& 0, \\ v_2 [m_2^2 + \lambda_2 v_2^2 + (\lambda_3 + \lambda_4) v_1^2] 
+ \mu_{12}^2 v_1 &=& 0.
\end{eqnarray}
Let $m_1^2 < 0$, $m_2^2 > 0$, and $|\mu_{12}^2| << m_2^2$, then
\begin{equation}
v_1^2 \simeq -{m_1^2 \over \lambda_1}, ~~~ v_2^2 \simeq {-\mu_{12}^2 v_1 \over 
m_2^2 + (\lambda_3 + \lambda_4) v_1^2}.
\end{equation}
Since the $\mu_{12}^2$ term breaks the $Z_3$ symmetry, it is natural \cite{th} 
for it to be small compared to $m_2^2$.  Thus $v_2 << v_1$ is obtained and 
since the first-generation quark and lepton masses are proportional to $v_2$, 
they are naturally small in this model.

The quark and lepton mass matrices (${\cal M}_u, {\cal M}_d, {\cal M}_e$) 
in this model are of the form
\begin{equation}
{\cal M} = \left[ \begin{array} {c@{\quad}c@{\quad}c} f_{11} v_2 & f_{12} v_1 
& f_{13} v_1 \\ 0 & f_{22} v_1 & f_{23} v_1 \\ 0 & f_{32} v_1 & f_{33} v_1 
\end{array} \right].
\end{equation}
This means that flavor-changing neutral currents exist in the Higgs sector. 
However, they would be absent if $\Phi_2$ is replaced by $\Phi_1$ in Eq.~(4). 
Hence the flavor-changing interactions must be contained in the terms
\begin{equation}
f^u_{11} \overline {u}_L u_R \left( \overline {\phi^0_2} - {v_2 \over v_1} 
\overline {\phi^0_1} \right) + f^d_{11} \overline {d}_L d_R \left( 
\phi^0_2 - {v_2 \over v_1} \phi^0_1 \right) + f^e_{11} \overline {e}_L e_R 
\left( \phi^0_2 - {v_2 \over v_1} \phi^0_1 \right) + h.c.,
\end{equation}
where $u_{L,R}$, etc. are in the basis of Eq.~(10) and are not themselves 
mass eigenstates.  The resultant off-diagonal terms are suppressed by the 
corresponding mixing angles and since $\phi^0_2$ is heavy and the coupling of 
$\phi^0_1$ is suppressed by $v_2/v_1$, the effect of flavor-changing neutral 
currents is very small in this model.

For illustration, consider the case of diagonal ${\cal M}_d$, then there are 
no tree-level flavor-changing interactions in the $down$ sector, but since 
${\cal M}_u$ must be rotated by the charged-current mixing matrix $(V_{CKM})$, 
flavor-changing interactions are unavoidable in the $up$ sector.  Allowing 
for the freedom to redefine $c_R$ and $t_R$, we can set $f_{32}=0$ for 
${\cal M}_u$ in Eq.~(10).  Then
\begin{equation}
m_u \simeq f_{11}^u v_2, ~~~ m_c \simeq f_{22}^u v_1, ~~~ m_t \simeq f_{33}^u 
v_1,
\end{equation}
and the $\overline {u}_L u_R$ term in Eq.~(11) becomes approximately 
\cite{small}
\begin{equation}
f_{11}^u (V^*_{ud} \overline {u}_L + V^*_{us} \overline {c}_L + V^*_{ub} 
\overline {t}_L) \left( V_{ud} u_R + {m_u \over m_c} V_{us} c_R + 
{m_u \over m_t} V_{ub} t_R \right) \left( \overline {\phi^0_2} - 
{v_2 \over v_1} \overline {\phi^0_1} \right) + h.c.
\end{equation}
in the mass-eigenvalue basis.  This conributes to $D^0 - \overline {D^0}$ 
mixing, i.e.
\begin{equation}
{\Delta m_{D^0} \over m_{D^0}} \simeq {B_D f_D^2 m_u^3 \over 3 m_2^2 v_2^2 
m_c} |V_{ud}^* V_{us}|^2.
\end{equation}
Using $f_D = 150$ MeV, $B_D = 0.8$, $m_u = 4$ MeV, $m_c = 1.25$ GeV, 
$|V_{ud}| \simeq 1$, $|V_{us}| \simeq 0.22$, and the experimental upper 
bound of $2.5 \times 10^{-14}$ \cite{pdg}, we find
\begin{equation}
m_2 v_2 > 24.4 ~{\rm GeV}^2,
\end{equation}
which may be satisfied for example with $m_2 = 1$ TeV and $v_2 = 25$ MeV. 
Another contribution to $D^0 - \overline {D^0}$ mixing is through 
$\Delta_u$ exchange, i.e.
\begin{equation}
{\Delta m_{D^0} \over m_{D^0}} \simeq {B_D f_D^2 h_u^2 m_u^2 \over 3 
m^2_{\Delta_u} m_c^2} |V^*_{ud} V_{us}|^2 \simeq 3 \times 10^{-15},
\end{equation}
for $h_u = 1$ and $m_{\Delta_u} = 1$ TeV, which is well below the present 
experimental upper bound.

At the TeV energy scale, the new scalars ($\Delta_u, \Delta_d, \Delta_e$, 
and $\Phi_2$) of this model are expected to be produced in abundance, 
especially $\Delta_u$ and $\Delta_d$ at the LHC (Large Hadron Collider). 
The decays of $\Delta_{u,d}$ into two jets should provide a clear signature 
for their discovery.  For the production of $\Delta_e$, the best accelerator 
would be an $e^- e^-$ collider.  It may also be produced in pairs at twice 
the energy at an $e^+ e^-$ collider, i.e. $e^+ e^- \to \Delta_e^* \Delta_e$. 
Of course, the $e^+ e^- \to e^+ e^-$ cross section is also modified through 
its exchange.

In conclusion, new physics at the TeV scale may be responsible for a 
measurable Majorana mass of the electron neutrino without requiring the 
near mass-degeneracy of all three neutrinos.  We have proposed a simple 
specific model of scalar diquarks and dileptons which allow this to happen, 
and is consistent with the present experimental bounds on flavor-changing 
neutral currents and other rare processes.

This work was supported in part by the U.~S.~Department of Energy
under Grant No.~DE-FG03-94ER40837.

\newpage
\bibliographystyle{unsrt}

\end{document}